\documentclass[12pt]{iopart}
\usepackage[utf8]{inputenc}
\usepackage{graphicx}
\usepackage{siunitx} 
\usepackage{xspace}
\usepackage{hyperref}
\usepackage{braket}
\usepackage{amsfonts}
\usepackage{bm}
\newcommand{\BeIon}{$^9$Be$^+$\xspace}

\begin{document}

\title{Multilayer ion trap technology for scalable quantum computing and quantum simulation}

\author{A. Bautista-Salvador$^{1,2,3}$, G. Zarantonello$^{1,2}$, H. Hahn$^{1,2}$, A. Preciado-Grijalva$^{1}$, J. Morgner$^{1,2}$, M. Wahnschaffe$^{1,2,3}$, C. Ospelkaus$^{1,2,3}$}    
\address{$^{1}$~Physikalisch-Technische Bundesanstalt, Bundesallee 100, 38116 Braunschweig, Germany}
\address{$^{2}$~Institute of Quantum Optics, Leibniz Universität Hannover, Welfengarten 1, 30167 Hannover, Germany}
\address{$^{3}$~Laboratory for Nano- and Quantum Engineering, Leibniz Universität Hannover, Schneiderberg 39, 30167 Hannover, Germany}

\ead{amado.bautista@ptb.de}

\vspace{10pt}
\begin{indented}
\item[] \today 
\end{indented}

\begin{abstract} 
We present a novel ion trap fabrication method enabling the realization of multilayer ion traps scalable to an in principle arbitrary number of metal-dielectric levels. We benchmark our method by fabricating a multilayer ion trap with integrated three-dimensional microwave circuitry. We demonstrate ion trapping and microwave control of the hyperfine states of a laser cooled \BeIon ion held at a distance of \SI{35}{\micro\meter} above the trap surface. This method can be used to implement large-scale ion trap arrays for scalable quantum information processing and quantum simulation.
\end{abstract}

\section{Introduction}\label{sec::intro}
Trapped ions are not only one of the most promising platforms for the practical implementation of quantum computing and quantum simulations, but also sensitive systems for measuring very small magnetic and electric fields. Typically, they are held in Paul or Penning traps at high vacuum, laser cooled close to absolute zero temperature, and their internal states coupled to their motion can be manipulated  with high fidelity by either laser fields~\cite{gaebler_high-fidelity_2016,ballance_high-fidelity_2016} or microwave radiation~\cite{harty_high-fidelity_2016, weidt_trapped-ion_2016}. However, scaling these elementary demonstrations to larger systems remains a formidable technological challenge~\cite{monroe_scaling_2013}.
 
Surface-electrode ion traps~\cite{seidelin_microfabricated_2006} represent a strong candidate for the realization of a quantum charge-coupled device (QCCD) \cite{wineland_experimental_1998,kielpinski_architecture_2002} for scaling quantum logic operations. Such an ion trap array could feature dedicated zones for storing, manipulation and read-out, thus promising a modular hardware for quantum computation and quantum simulation~\cite{lekitsch_blueprint_2017}. Conventionally, in surface-electrode ion traps all electrodes are built in a single plane by standard microfabrication techniques~\cite{hughes_microfabricated_2011}. First integration of key scalable elements into a single layer chip such as micro-optical components~\cite{merrill_demonstration_2011}, nanophotonic waveguide devices~\cite{mehta_integrated_2016} or microwave conductors~\cite{ospelkaus_microwave_2011} have been demonstrated. However, interconnecting separated components built in this system imposes new challenges on trap design where signal lines have to be routed around other elements. Therefore, the realization of a highly integrated large-scale ion trap device requires a more flexible approach where signal routings can be distributed on vertically well-separated levels of interconnects. 

Demonstrations of multilayer processes in ion traps so far are based on techniques borrowed from  MicroElectroMechanical Systems (MEMS)~\cite{cho_review_2015,hong_guidelines_2016} or CMOS~\cite{stick_demonstration_2010,maunz_high_2016}; however the resulting trap structures are limited to thin interconnect levels. Moreover, there is a need of a nearly material-independent processing capable of including most dielectric substrates and thick metallization. Any fabrication process will have to comply with the specific requirements of an ion trap, such as a material mix which features extremely low material outgassing and needs to be compatible with ultra-high vacuum (UHV) operation, low dielectric losses and non-magnetic metal surfaces. Often one is concerned about shielding the ion(s) from patch potentials due to exposed dielectrics. Specially, for the top-level the electrode interspacings should have at least a width equal to its height~\cite{schmied_electrostatics_2010}, or in other words electrode gaps of an aspect ratio higher than 1:1.  

Here we present a robust fabrication method, scalable to an in principle arbitrary number of planarized thick metal-dielectric layers, enabling the realization of scalable ion trap devices. The method complies with the stringent requirements of a scalable ion trapping array, allowing the fabrication of complex trap designs using relatively forgiving fabrication techniques on nearly any type of substrate. To demonstrate the approach, we fabricate and operate a multilayer ion trap chip with three-dimensional (3-D) microwave circuitry towards the realization of high fidelity muli-qubit gates~\cite{ospelkaus_microwave_2011,timoney_quantum_2011}.  
 
\section{Fabrication Methods}\label{sec::FabricationMethods}
Methods for building surface-electrode ion traps~\cite{seidelin_microfabricated_2006} or atom chips~\cite{treutlein_coherent_2008,fortagh_magnetic_2007} are typically based on standard semiconductor processing. For the simplest case, in which all metal electrodes are aligned in a single plane, a generic fabrication workflow consists of a three-step processing: wafer patterning, electrode formation and electric insulation. Depending on the requirements one will choose between different materials and processing methods at hand. In what follows we will describe our own fabrication methods to build single layer and multilayer microfabricated ion traps.

\subsection{Single Level Processing (SLP) Method}
\begin{figure}[!h] 
	\centering
	\includegraphics[width=\columnwidth]{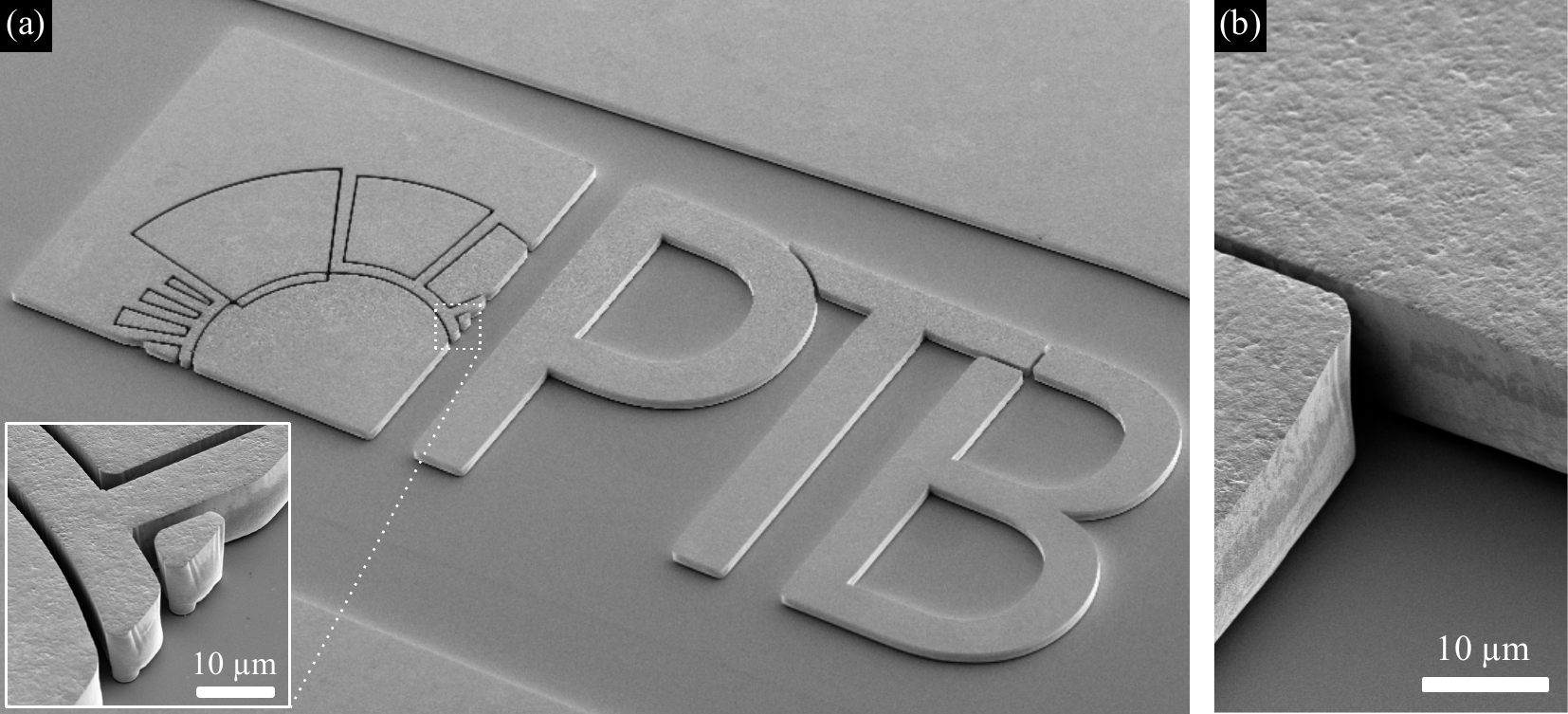}
	\caption{Gold structures with high aspect ratio supported on sapphire. (a) Au structures separated by gaps of about \SI{2}{\micro\meter} with a high aspect ratio of 5:1 (inset). (b) Pair of Au electrodes separated by a \SI{0.8}{\micro\m} gap, resulting in a gap with an aspect ratio of 14:1. For both SEM micrographs shown in (a) and (b) the sample is tilted by \SI{45}{\degree}.}
	\label{fig::SEM::fig1} 
\end{figure}

For the Single Level Processing (SLP) method all steps are carried out on 3-inch-diameter wafers in a fabrication line located at Physikalisch-Technische Bundesanstalt (PTB), Braunschweig. We have fabricated similar structures to the ones presented in Fig.~\ref{fig::SEM::fig1} on AlN, sapphire, organic polymers and high resistivity (HiR) float zone (FZ) Si wafers, demonstrating the compatibility of the method with a wide range of substrates suitable for ion trap technology. 

The first step during wafer preparation involves the deposition of a Ti adhesion layer (\SI{15}{\nano\meter} thin) and an Au seed layer (\SI{50}{\nano\meter} thin) on top of the substrate by resistive evaporation. The first film acts as an adhesion promoter between the substrate and the Au seed layer, and the second film serves as a starting conductive layer for a later electrodeposition step.  

Second, to define the trap geometry, a 25-\SI{}{\micro\meter}-thick positive or 16-\SI{}{\micro\meter}-thick negative resist is spin coated on top of the Au seed layer and the wafer is exposed to UV light by contact lithography. A subsequent development of the exposed resist results in open areas on the substrate which are filled to a desired thickness by electrodeposition of Au in a sulphite-based bath. 

Finally, after gold electroplating the resist mask is removed chemically and the wafer is cleaned under oxygen-based plasma etching. Additionally, the wafer is exposed to a fluorine-based plasma to further remove possible resist debris. Immediately afterwards the seed Au layer is removed via Ar etching and the Ti layer removed by a fluorine-based plasma etching. 

This method allows the fabrication of gold structures with high aspect ratios as exemplified in Fig~\ref{fig::SEM::fig1}. Gold structures with a width as narrow as \SI{5}{\micro\meter} and gap separation as narrow as \SI{2}{\micro\meter} are shown in Fig.~\ref{fig::SEM::fig1}(a). Another example is depicted in Fig.~\ref{fig::SEM::fig1}(b) consisting of a pair of gold electrodes separated by a gap with an aspect ratio of 14:1. One additional advantage of the method is that after dry etching of the Au/Ti bilayer the resulting trap surfaces have a superior finishing quality compared to the commonly used wet etching.

\subsection{Multilevel Processing (MLP) Method}\label{sec::MLP_section}
In this section a Multilevel Processing (MLP) method is presented, which combines techniques borrowed from MicroElectroMechanical Systems (MEMS) and Integrated Circuits (IC) processing. The method is also compatible with other common substrates used for ion trap technology such as silicon, sapphire, borosilicate glass and quartz. 

To demonstrate the simplicity and robustness of our method we have fabricated an ion trap with integrated (3-D) three-dimensional microwave circuitry. It comprises a lower interconnect level $\text{L}_1$ and an upper electrode level $\text{L}_2$.  An additional vertical interconnect access $\text{V}_1$, called via, allows microwave signals to be transmitted between levels. A more detailed description of the microwave and quantum logic aspects of the trap design and the corresponding characterization will be covered elsewhere~\cite{hahn_multilayer_nodate}.

\begin{figure}[!h]
	\centering
	\includegraphics[width=\columnwidth]{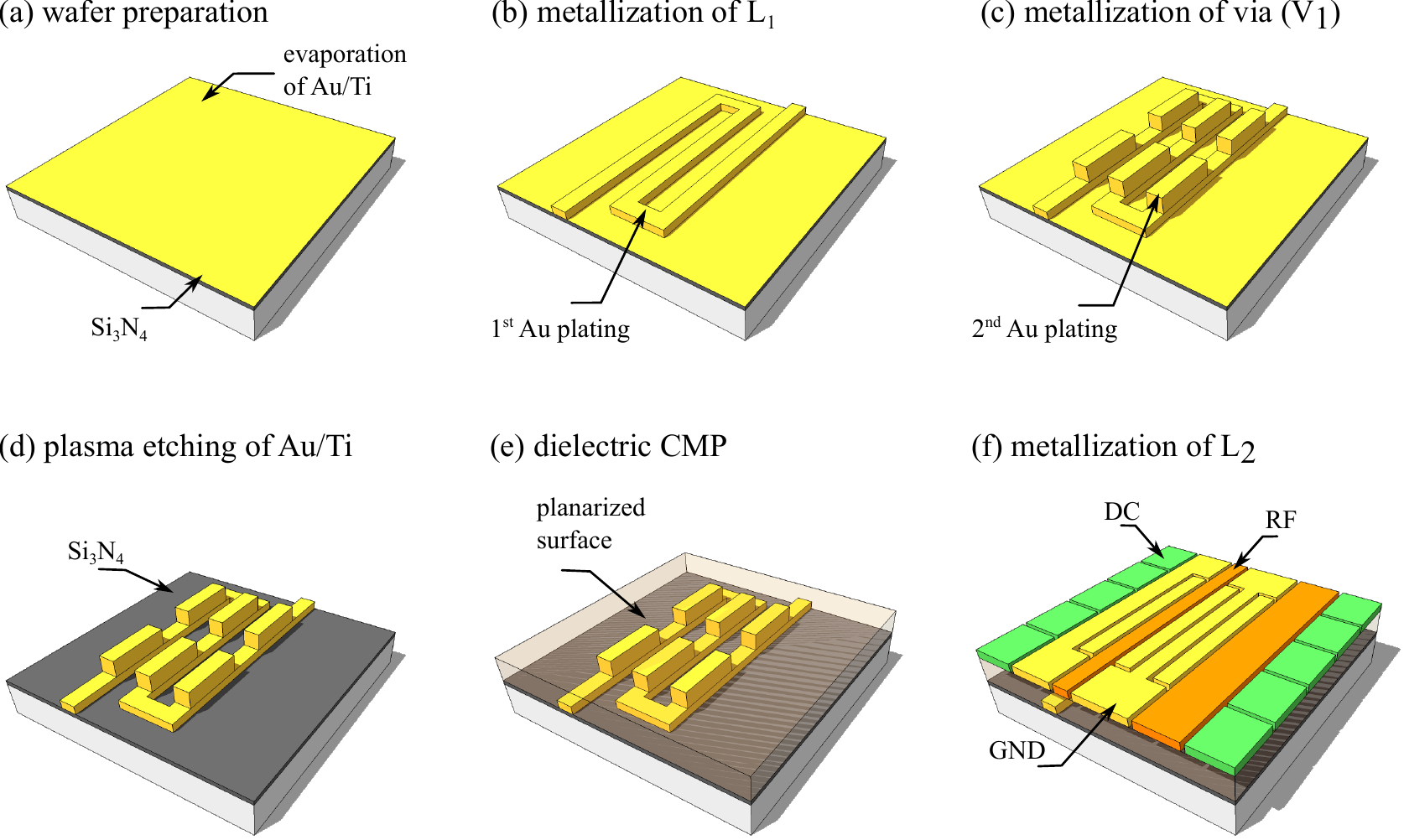}
	\caption{Schematics of the multilevel processing flow. (a) A wafer is coated with a conductive Au/Ti bilayer film. (b) A lower level ($\text{L}_1$) is formed via electroplating. (c) a via V$_1$ is electroplated on top of $\text{L}_1$.  (d) Dry etching of Au/Ti bilayer film. (e) A polymer is spin coated over both L$_1$ and V$_1$ and subsequently planarized by CMP. (f) A second metallization L$_2$ repeating steps (a), (b) and (d) on a planarized dielectric surface. Exposed dielectric between electrodes in L$_2$ is removed by plasma etching down to the top surface of the Si$_3$N$_4$ layer.} %change pic in (f)
	\label{fig::MLFlow:fig2} 
\end{figure}

The method presented here mainly consists of six processing steps: (a) wafer preparation, (b) metallization of lower level $\text{L}_1$, (c) metallization of via $\text{V}_1$, (d) removal of seed layer, (e) deposition and planarization of dielectric layer $\text{D}_1$ and (f) metallization of upper level $\text{L}_2$. A schematics of the fabrication flow is given in Fig.~\ref{fig::MLFlow:fig2}.  
 
The supporting material is a 3-inch silicon wafer with high resistivity ($\sigma>\SI{1e4}{\ohm\cm}$). On top of it and as shown in Fig.~\ref{fig::MLFlow:fig2}(a), a 2-\SI{}{\micro\m}-thick film of Si$_3$N$_4$ is deposited by physical enhanced chemical vapor deposition (PECVD). This dielectric film may improve trap operation by avoiding detrimental diffussion of Au into silicon and increasing the flashover voltage as demonstrated in Ref.~\cite{sterling_increased_2013}. Thereafter, a 10-\SI{}{\nano\m}-thin layer of Ti and a 50-\SI{}{\nano\m}-thin layer of Au are thermally evaporated on top of Si$_3$N$_4$. 

To build the lower level $\text{L}_1$ on top of Si$_3$N$_4$/Si, a negative resist is spin coated and patterned via UV lithography. Once the negative resist is developed to form a resist mold, gold electrodes are grown by electroplating as depicted in Fig.~\ref{fig::MLFlow:fig2}(b). After electroplating is completed, the resist mask is stripped and the wafer cleaned under plasma etching.

For the metallization of the via $\text{V}_1$ we repeat the photolithography and electroplating steps presented in (b) but this time on top of $\text{L}_1$ by using a thick ($>\SI{12}{\micro\meter}$) developed negative resist as a plating mold. Electroplated vias on top of L$_1$ are depicted in Fig.~\ref{fig::MLFlow:fig2}(c) after stripping the negative resist mask and plasma cleaning of the wafer.

To remove the Au seed layer and the Ti adhesion layer we use the last dry etching step from the SLP method. This step allows a controllable etch of Au and Ti of \SI{50}{\nano\m\per\minute} and \SI{10}{\nano\m\per\minute} respectively, resulting in a minimal change of the surface quality on top of both $\text{L}_1$ and $\text{V}_1$ surfaces. The electrically isolated elements on $\text{V}_1$ and $\text{L}_1$ are schematically illustrated in Fig.~\ref{fig::MLFlow:fig2}(d).  

A dielectric layer is then spin coated on top of $\text{L}_1$ and $\text{V}_1$ and thermally cured (Fig.~\ref{fig::MLFlow:fig2}(e)). After thermal curing, excess material is present on top of the underlaying structures in $\text{L}_1$ and $\text{V}_1$. The imprinted dielectric topography is globally planarized through a chemical-mechanical polishing (CMP) step, which is stopped at the top of $\text{V}_1$ or close to it. To assure electrical contact between $\text{V}_1$ and the subsequent level $\text{L}_2$ a local etch-back process is performed. 

To define the top metal layer the SLP method is again employed but this time on top of the planarized polymer surface (Fig.~\ref{fig::MLFlow:fig2}(f)). Once the plating has been completed and the resist mold removed, the remaining polymer film between gaps underneath L$_2$ is etched down to the Si$_3$N$_4$ layer by a fluorine-based plasma to hide possible patch potentials built on the exposed insulator.
 
\section{Fabrication outcome and trap operation}\label{sec::IonTrapping}
Here we briefly present the design and characterization of a trap with 3-D microwave conductors integrated into a microfabricated ion trap using the MLP method. The microwave circuitry is embedded to implement quantum logic operations using near-field microwaves \cite{ospelkaus_trapped-ion_2008,ospelkaus_microwave_2011,warring_techniques_2013,carsjens_surface-electrode_2014}. The specific design is discussed elsewhere in detail \cite{hahn_multilayer_2018} and here only described as one of many scenarios that benefits from the multilayer technology. 

\begin{figure}[!h]
	\centering
	\includegraphics[width=\columnwidth]{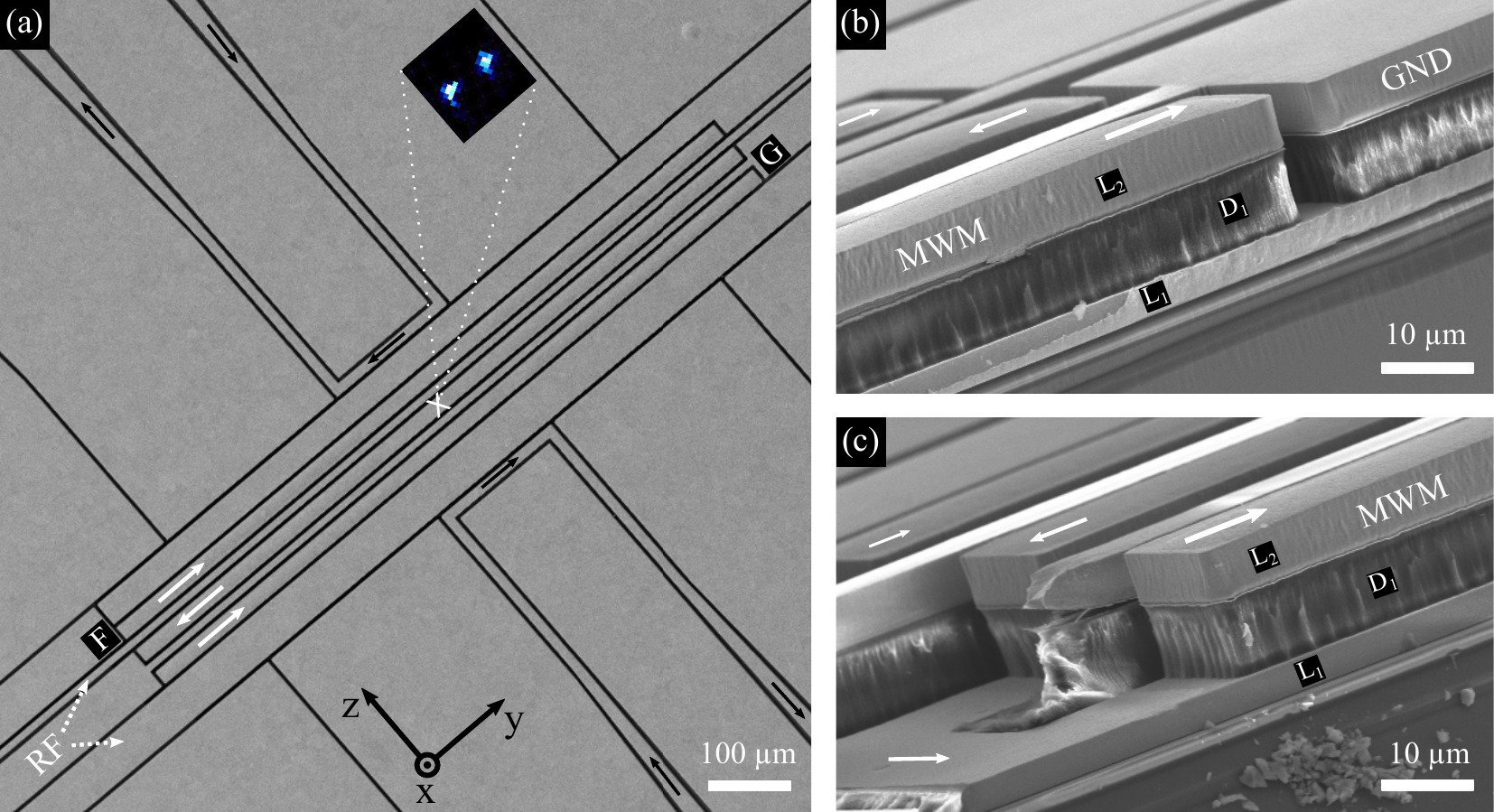}
	\caption{SEM micrograph of the central part of a trap chip fabricated using the MLP method. White and black arrows are drawn on top of the microwave conductors as a guide for the eye. (a) upper metal level L$_2$ of the trap chip in which black lines correspond to electrode gaps of about \SI{5}{\micro\meter}. Ions are trapped at $d=\SI{35}{\micro\meter}$ above the position marked as ``X''. (b) Part of a cleaved chip showing a cross section mainly on the $xy-$plane around position ``G''. (c) Similar to (b) but with a close-up corresponding to position ``F'' in (a).}
	\label{fig::SEM:fig3} 
\end{figure}

In the upper level (L$_2$) the trap includes two RF electrodes and ten DC electrodes to confine the ions to a local minimum ($x_0,z_0$), see Fig.~\ref{fig::SEM:fig3}(a). A microwave signal (white arrows) of frequency \SI{1}{\giga\hertz} can be applied on a 3-D microwave meander (MWM) conductor between two contact points labeled as ``F'' and ``G'', thus generating an oscillating magnetic near-field gradient $B'$ in the $xz-$radial plane with a local minimum at $(x_1,z_1)$. The three apparent independent microwave conductors indicated by the white arrows are indeed part of a single 3-D microwave meander connected to L$_1$ by vias and routed over L$_1$ as inidicated in Fig.~\ref{fig::SEM:fig3}. A central part of a diced trap chip ($\SI{5}{\mm}\times\SI{5}{\mm}$) fabricated using the MLP method is presented in Fig.~\ref{fig::SEM:fig3}(a). There are also two additional microwave conductors (MWC) surrounding the central DC electrodes, in which an oscillating current (black arrows) can be applied to produce an oscillating $B$ field.  

A cleaved chip revealing a cross-section view of the metal-dielectric stack around position ``F'' position ``G'' is shown in Fig.~\ref{fig::SEM:fig3}(b) and Fig.~\ref{fig::SEM:fig3}(c)), respectively. The ion trap (RF and DC electrodes) as well as the uppermost part of the MWM conductor are entirely located in L$_2$, whereas the microwave signals and the ground plane are routed between L$_1$ and L$_2$ through vias in V$_1$ (not visible in the micrograph but behind D$_1$ in Fig.~\ref{fig::SEM:fig3}(b) and (c)). 

Removing both Au and Ti films by means of dry etching has improved the trap surface quality. For a similar chip as the one here presented an rms roughness $R_{rms}= \SI{8.3\pm0.5}{\nano\meter}$ is obtained by atomic-force microscopy over an area of $\SI{25}{\micro\meter}\times\SI{25}{\micro\meter}$. This represents a two-order of magnitude improvement when compared to a wet etching process using aqua regia~\cite{wahnschaffe_single-ion_2017}. These nearly mirror-like surfaces are relevant since there is a reduction of stray light scattered in the direction perpendicular to the trap surface during resonance fluorescence imaging for ion state detection. Also an ion trap with minimal surface roughness might be less prone to anomalous motional heating at cryogenic temperatures~\cite{lin_effects_2016}. 
 
The diced trap chip is glued onto a copper block and wirebonded to a custom printed circuit board for filtering and signal routing. The whole assembly is installed in a vacuum system at a pressure better than \SI{1e-11}{\milli\bar} and connected to an in-vacuum coaxial resonator similar to the one used in Ref.~\cite{jefferts_coaxial-resonator-driven_1995}. For ion loading we employ a laser ablation scheme~\cite{wahnschaffe_single-ion_2017} and subsequent two-photon ionization using \SI{235}{\nano\meter} light~\cite{leibrandt_laser_2007}. Single \BeIon ions are loaded at \SI{35}{\micro\meter} above the upper surface of $\text{L}_2$ around the position ``X'' (see Fig.~\ref{fig::SEM:fig3}(a)). 

\begin{figure}[!h]
	\centering
	\includegraphics[width=\columnwidth]{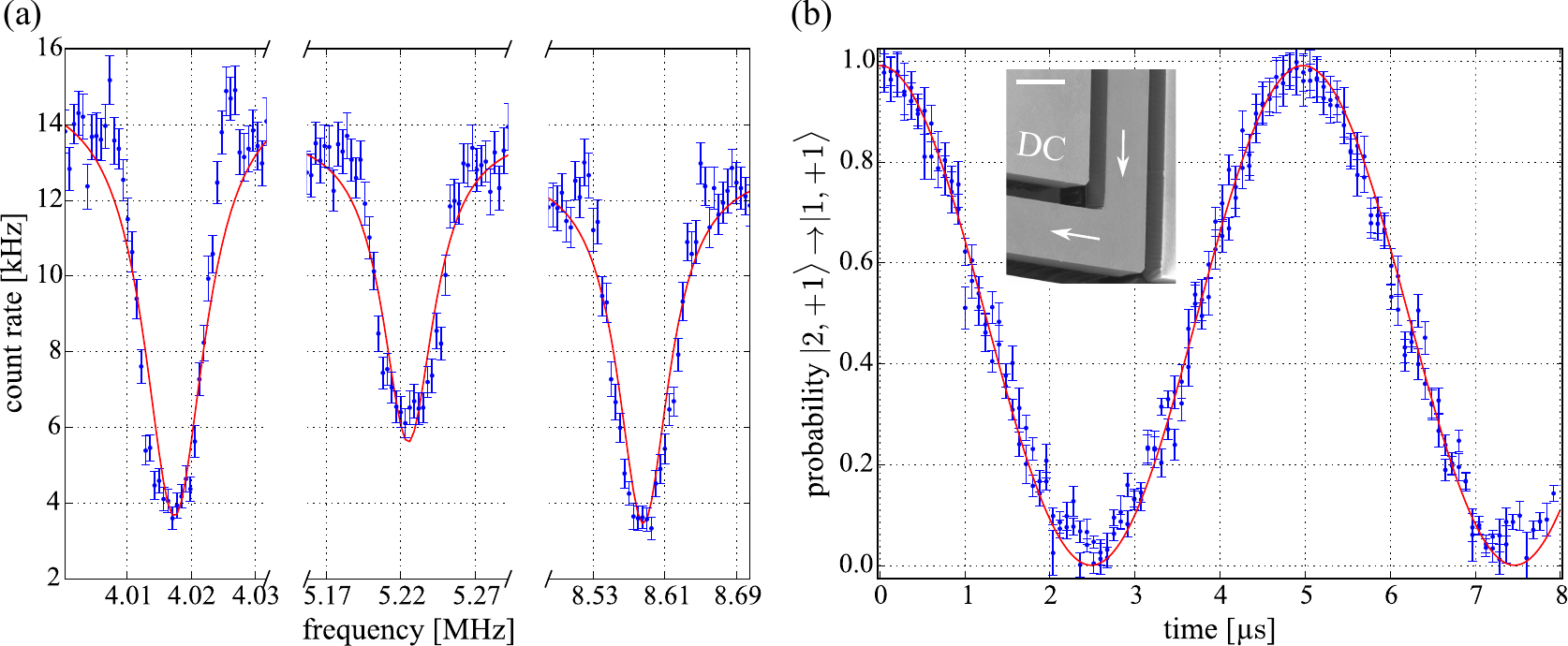}
	\caption{Secular frequencies and microwave qubit transition. (a) Motional frequency measurements using the DC tickle method. The (red) solid lines show Lorentzian fits to the data. The left axis corresponds to the photon average counts after a fluoresence detection of \SI{1.3}{\milli\second}. (b) Rabi flopping on the qubit transition using a microwave conductor embedded in the trap surrounding a central DC electrode (inset). The white scale bar corresponds to \SI{10}{\micro\meter} and arrows are used only as a guide for the eye.} 
	\label{fig::ModRabi:fig4} 
\end{figure}

We supply to the trap an RF drive frequency of $\Omega_{RF} = 2\pi\times\SI{176.5}{\MHz}$ with amplitude $\text{V}_{RF}=\SI{100}{\volt}$ and DC voltages ranging within $\pm\SI{25}{\volt}$. To determine the trap frequencies we apply an oscillating tickle voltage to one of the DC electrodes and scan the frequency~\cite{home_normal_2011}. Once the tickle drive is resonant with a secular frequency, the motion of the ion is excited and the ion fluorescence drops (Fig.~\ref{fig::ModRabi:fig4}(a)). We measure secular trap frequencies of $(\omega_\mathrm{y},\omega_\mathrm{LF}, \omega_\mathrm{HF}) = 2\pi\cdot(4.02,5.23,8.59)\, \text{MHz}$, where the high-frequency (HF) and low-frequency (LF) radial modes form an angle of $-5.9^\circ$ relative to the $x$-axis and $z$-axis, respectively.

Finally we employ the integrated microwave conductors to manipulate the internal state of the ion. Fig.~\ref{fig::ModRabi:fig4}(b) shows Rabi oscillations on the qubit transition $\ket{F=2,\,m_F=+1}\,\rightarrow\ket{F=1,\,m_F=+1}$~\cite{wahnschaffe_single-ion_2017} of the electronic ground state $^{2}S_{1/2}$ of a single \BeIon ion at an external magnetic field of $|\bm{B}_{0}|=\SI{22.3}{\milli\tesla}$ when applying a microwave current of frequency $\omega_0\simeq\SI{1082.55}{\mega\hertz}$ to one of the MWC conductors. Here, $F$ is referring to the total angular momentum $\bm{F}$ and $m_F$ the quantum number of its projection on $\bm{B}_{0}$.  The state readout is carried out via ion fluorescence detection on the closed-cycling transition $\ket{S_{1/2},F=2,\,m_F=2}\rightarrow\ket{P_{3/2},m_J=+3/2,m_I=+3/2}$, combined with suitable microwave transfer pulses~\cite{wahnschaffe_single-ion_2017}. 

\section{Conclusion and Outlook}\label{sec::conclusions}
We have presented a novel multilayer method for fabricating scalable surface-electrode ion traps. The flexibility and robustness of the method allows to benchmark the integration of 3-D microwave circuitry into a multilayer ion trap. Furthermore, we have demonstrated successful trapping of \BeIon and basic qubit manipulation by applying microwave oscillating currents on one of the conductors. 

The MLP method presented here can in principle be extended to a nearly arbitrary number of layers to comply with the stringent needs of scaling surface-electrode ion traps. Moreover, the method permits the integration of three-dimensional and planarized features with high aspect ratio. This technique opens new routes towards the realization of more complex and powerful ion trap devices.

In contrast to a typical CMOS situation where the ``device'' is fabricated on top of the substrate, with interconnect layers on top of the device, in our case the ``device'' is the top electrode layer which is controlling the ion(s), whereas the layers closer to the substrate are used as interconnects. In the future, these lower interconnects may be combined with through-wafer vias to achieve contacting of the ion trap chip from the backside, eliminating the need of wirebonds and likely obstruction of laser beams. Through-wafer slots for back-side ion loading can also be produced in the same way. These same techniques could be applied to realize so-called analog quantum simulators in ion trap arrays \cite{porras_effective_2004,mielenz_arrays_2016,bruzewicz_scalable_2016}, possibly with integrated control \cite{chiaverini_laserless_2008}. Moreover, such an approach may enable the embedding of complex integrated components such as trench capacitors~\cite{allcock_heating_2012, guise_ball-grid_2015}, low-loss integrated waveguides~\cite{west_low_2018}; or the realization of more elaborate devices including reliable ion-transport junctions~\cite{moehring_design_2011,wright_reliable_2013}, increased optical access \cite{maunz_high_2016} or manipulation of scalable arrays of two-dimensional trapped ion systems~\cite{mielenz_arrays_2016,bruzewicz_scalable_2016}.

The MLP method can also be used to extend multilayer ``atom chips''~\cite{fortagh_magnetic_2007,trinker_multilayer_2008,bohi_coherent_2009} or to fabricate scalable hybrid atom-ion traps~\cite{harter_cold_2014,bahrami_operation_2018} for quantum many-body physics experiments and quantum sensing with neutral atoms. In this context the thick metal conductors can support substantial currents required for magnetic trapping and the planarization together with the demonstrated minimized surface roughness allows the implementation of mirror-like surfaces and transfer coatings for integrated magneto-optical traps.

\section*{Acknowledgements}
We acknowledge support by the PTB cleanroom facility team, in particular T. Weimann and P. Hinze. We also acknowledge support by the LNQE cleanroom staff in particular O. Kerker. We acknowledge funding from PTB, QUEST, LUH, NTH (project number 2.2.11) and DFG through CRC 1227 DQ-\textit{mat}, project A01. 
  
\section*{References}
\bibliography{qc}
\bibliographystyle{iopart-num}
\end{document}